\documentstyle[version2,preprint,aps]{revtex} 
\begin{document}
\preprint{MAD/TH/93-9}
\draft 
\begin{title} 
Casimir operators of the exceptional group $F_4$: the chain 
$B_4\subset F_4\subset D_{13}$
\end{title} 
\author{Adam M. Bincer} 
\begin{instit} 
Department
of Physics, University of Wisconsin--Madison,\\ 
Madison, Wisconsin
53706 
\end{instit} 
\begin{abstract} 
Expressions are given for the Casimir operators of the exceptional 
group $F_4$ in a concise form similar to that used for the 
classical groups. The chain $B_4\subset F_4\subset D_{13}$ is used to 
label the generators of $F_4$ in terms of the adjoint and spinor 
representations of $B_4$ and to express the 26-dimensional 
representation of $F_4$ in terms of the defining representation of 
$D_{13}$. Casimir operators of any degree are obtained and it is 
shown that a basis consists of the operators of degree 2, 6, 8 and 
12.
\end{abstract}

\section{INTRODUCTION}

Although a general formula exists for the quadratic Casimir
operator for any group this is not the case for operators of
higher degree. Efficient expressions have been developed over the
years for all the Casimir operators of the classical groups, but
not for the exceptional groups. Berdjis\cite{berdjis} gives the
desired Casimir operators implicitly. Until recently explicit
results were available only for $G_2$. The degree 6 Casimir of
$G_2$ was given in the work of Hughes and Van der
Jeugt\cite{hughes} by an expression involving 29 terms and in the
work by Bincer and Riesselmann\cite{bin_ries} by an expression
involving 23 terms. These results were obtained using computers
and leave something to be desired.

Quite recently I have developed a different approach and obtained
for $G_2$ results very much alike to those for the classical
groups\cite{classical}. Moreover it would seem that the same
approach should work for the other exceptional groups. In the
present work I address the group $F_4$ and leave the $E_{6,7,8}$
for a future paper.

This paper is organized as follows. In the next Sec.\ after
explaining the use of the chain $B_4\subset F_4\subset D_{13}$ I
obtain concise expressions for the Casimir operators of $F_4$.
These require the knowledge of the generators of $D_{13}$
projected into $F_4$. To obtain this projection I describe in the
next Sec.\ the 26-dimensional representation of $F_4$ and then
obtain in the following Sec.\ the desired projection. In the
Conclusion I discuss the quadratic Casimir operator of $F_4$ and
demonstrate that the independent Casimir operators are of degree
2, 6, 8 and 12 (corresponding to the exponents of $F_4$ being 1,
5, 7 and 11).

\section{The Casimir operators of $D_{13}$ and $F_4$}

My approach makes use of the chain $B_4\subset F_4\subset D_{13}$.
The subgroup $B_4$ of $F_4$ is used to label the generators of
$F_4$. $F_4$ is embedded in $D_{13}$ because the
smallest-dimensional representation of $F_4$ is 26-dimensional and
orthogonal and $D_{13}$ is the orthogonal group in 26 dimensions.

I denote the 36 generators of $B_4$ as $B_{\alpha
}^\beta=-B_{\bar\beta }^{\bar\alpha }$, with indices ranging from $-4$
to $+4$, zero {\em included\/}, $\bar\alpha \equiv-\alpha $. The
hermitian property is expressed in this basis as $B_{\alpha
}^{\beta \dagger}=B_{\beta }^{\alpha }$. I denote the generators
of $F_4$ as $B_{\alpha }^{\beta }$ and $S^{pqrs}$, corresponding
to the decomposition of the {\bf 52} (the adjoint) of $F_4$ into
the {\bf 36} and {\bf 16} of $B_4$, where the {\bf 36} is the
adjoint, i.e., the $B_{\alpha }^{\beta }$, and the {\bf 16} is the
spinor $S^{pqrs}=\left( S^{\overline{pqrs}} \right)^{\dagger}$,
$p,q,r,s=\pm$. The $B_4\subset F_4$ relation is exhibited in the
extended Dynkin diagram
\begin{center}

	\begin{tabular}{c@{\hspace*{18pt}}c@{\hspace*{18pt}}c@{\hspace*{29pt}}c
	@{\hspace*{35pt}}l}
		$\alpha _0$ & $\alpha _1$ & $\alpha _2$ & $\alpha _3$ & $\alpha
		_4$\\
		\multicolumn{5}{l}{
		\begin{picture}(225,10)(-20,-2.5)
			\put(0,5){\circle{5}}
			\put(2.5,5){\line(1,0){50}}
			\put(55,5){\circle{5}}
			\put(57.5,5){\line(1,0){50}}
			\put(110,5){\circle{5}}
			\put(112.5,7){\line(1,0){50}}
			\put(112.5,3){\line(1,0){50}}
			\put(165,5){\circle*{5}}
			\put(167.5,5){\line(1,0){50}}
			\put(220,5){\circle*{5}}
		\end{picture}}\\
		$u_1-u_2$ & $u_2-u_3$ & $u_3-u_4$ & $u_4$ & $-\frac{1}{2}\left(
		u_1+u_2+u_3+u_4 \right)$
	\end{tabular}
\end{center}
with $B_4$ obtained by omitting $\alpha _4$ and $F_4$ obtained by
omitting $\alpha _0$. That is to say: the $\alpha _i,\ 1\leq i\leq
4$, are the simple roots of $F_4$, while the $\alpha _j,\ 0\leq
j\leq 3$ are the simple roots of $B_4$. The information encoded
in the Dynkin digram  is made explicit by setting $\alpha
_0=u_1-u_2, \alpha _1=u_2-u_3,\alpha _2=u_3-u_4,\alpha
_3=u_4,\alpha _4=-\frac{1}{2}(u_1+u_2+u_3+u_4)$, where the $u_i$
are orthogonal unit vectors.

I denote the generators of $D_{13}$ as
$D_a^b=-D_{\bar b}^{\bar a},\left( D_a^b \right)^\dagger=D_b^a$, zero {\em
excluded\/}. The commutation relations of $D_{13}$ in this basis are
%
%
\begin{equation}
\label{eq1}
	\left[ D_{a}^{b},D_{c}^{d} \right]=\delta
	_{c}^{b}D_{a}^{d}-\delta _{a}^{d}D_{c}^{b}+\delta _{\bar
	b}^dD_{c}^{\bar a}-\delta _{c}^{\bar a}D_{\bar b}^d
\end{equation}
It follows from Eq.\ (\ref{eq1}) that
%
%
\begin{equation}
\label{eq2}
	\left[ D_{a}^{b},\left( D^k \right)_{c}^{d} \right]=\delta
	_{c}^{b}\left( D^k \right)_{a}^{d}-\delta _{a}^{d}\left( D^k
	\right)_{c}^{b}+\delta _{\bar b}^{d}\left( D^k \right)_{c}^{\bar
	a}-\delta _{c}^{\bar a}\left( D^k \right)_{\bar b}^d
\end{equation}
where I define the $k$th power, $k\geq 1$, by
%
%
\begin{equation}
\label{eq3}
	\left( D^k \right)_{a}^{b}=\left( D^{k-1} \right)_{a}^{c}D_{c}^{b}=
	D_{a}^{c}\left( D^{k-1} \right)_{c}^{b},\qquad \left( D^0
	\right)_{a}^{b}=\delta _{a}^{b}
\end{equation}
(summation convention understood). It now follows that if I define
%
%
\begin{equation}
\label{eq4}
	{\cal C}_k(D_{13})=(D^k)_{a}^{a}
\end{equation}
then these ${\cal C}_k$ commute with the generators of $D_{13}$
and so are Casimir operators of $D_{13}$ of degree $k$. Equation
(\ref{eq14}) provides an elegant expression for the Casimir
operators of $D_{13}$ and is an example of the type of expressions
valid for all the classical grups. All this is well-known and goes
back to Perelomov and Popov\cite{perelomov}. I remark that the 13
independent Casimirs of this type are of degree $k=2s,1\leq
s\leq13$. This is because it follows from the antisymmetry
property $D_{a}^b=-D_{\bar b}^{\bar a}$ that the Casimirs for
$k=\mbox{odd}$ can be expressed in terms of those for
$k=\mbox{even}$, and it follows from the Cayley-Hamilton theorem
that Casimirs of degree $k>26$ can be expressed in terms of those
for $k\leq26$. I note further that the Casimir operator of degree
26 can be expressed in terms of the square of a Casimir of degree
13 [which is not of the form given by Eq.\ (\ref{eq4})] and so the
integrity basis for the Casimirs contains those of degree
$k=2s,1\leq s\leq 12$, and $k=13$, which agrees with the fact that
the degrees $k$ of the Casimirs in the basis should be equal to
the exponents of $D_{13}$ plus one.

We next observe that under the restriction of $D_{13}$ to $F_4$
the adjoint representation of $D_{13}$ decomposes thus
%
%
\begin{equation}
\label{eq5}
	\bf 325=52+273
\end{equation}
where the $\bf 325$ refers to the adjoint of $D_{13}$ and the
${\bf 52}$ to the adjoint of $F_4$. Thus we can express the
generators $D_{a}^b$ of $D_{13}$ in terms of the generators of
$F_4$ and the components of the $\bf 273$-plet. We now obtain the
Casimir operators of $F_4$ by observing that they are given by
Eq.\ (\ref{eq4}) in which the $D_{a}^b$ are replaced by their
projections into $F_4$, i.e.,
%
%
\begin{equation}
\label{eq6}
	{\cal C}_k(F_4)=(\tilde D^k)_{a}^a
\end{equation}
where
%
%
\begin{equation}
\label{eq7}
	\tilde D_a^b=D_a^b|_{{\bf273}=0}
\end{equation}
I mean by Eq.\ (\ref{eq7}) that the projected $\tilde D_{a}^b$ are
given by expressing the $D_{a}^b $ in terms of the generators of
$F_4$ and members of the $\bf 273$-plet and then setting the
contribution of the $\bf 273$-plet equal to zero.

\section*{The 26-dimensional representation of $F_4$}

To obtain the projected $\tilde D$ I need to obtain first explicit
formulas for the 26-dimensional representation of $F_4$.

The generators $D_{a}^b$ of $D_{13}$ are given in the defining
26-dimensional representation as the following $26\times26 $
matrices:
%
%
\begin{equation}
\label{eq8}
	D_{a}^b=I_{ab}-I_{\overline{ba}}
\end{equation}
where $I_{ab}$ is the $26\times26$ matrix with matrix elements
%
%
\begin{equation}
\label{eq9}
	(I_{ab})_{jk}=\delta _{aj}\delta _{bk}
\end{equation}
with the labels $j,k$ taking on the same values as $a,b$: $-13\leq
j,k\leq13$, zero excluded.

The Cartan generators of $F_4$ are given in the 26-dimensional
representation by the $26\times26$ matrices as follows:
\begin{eqnarray}
	\label{eq10}
	h_1 &=& D_{5}^5+D_{6}^6-D_7^7+D_{8}^8-D_{9}^9-D_{10}^{10}\\
	\label{eq11}
	h_2 &=& D_{3}^3+D_{4}^4-D_{5}^5-D_{6}^6+D_{10}^{10}-D_{11}^{11}\\
	\label{eq12}
	h_3 &=& \frac{1}{2}\left(
D_{2}^2-2D_{3}^3-D_{4}^4+D_{6}^6-D_{8}^8+D_{9}^9-D_{10}^{10}+D_{11}^{11}-D_{12}^{12}
	\right)\\
	\label{eq13}
	h_4 &=& \frac{1}{2}\left(
-2D_{2}^2+D_{3}^3-D_{4}^4+D_{5}^5-D_{6}^6+D_{7}^7-D_{9}^9+D_{12}^{12}-D_{13}^{13}
	\right)
\end{eqnarray}
These are precisely the same expressions as were obtained by
Patera\cite{patera} and Ekins and Cornwell\cite{ekins} if I
relabel their rows and columns thus: their
$1\rightarrow\mbox{mine}-13$, their
$2\rightarrow\mbox{mine}-12,\ldots$, their
$13\rightarrow\mbox{mine}-1$, their
$14\rightarrow\mbox{mine}+1,\ldots$, their
$26\rightarrow\mbox{mine}+13$.

Given these explicit matrices for the Cartan generators $h_i$, the
associated generators $e_i$ and $f_i=e_i^\dagger$ in the Chevalley
basis are found from the equations\cite{ekins}
%
%
\begin{equation}
\label{eq14}
	\left[ e_j,h_k \right]=A_{kj}e_j,\qquad \left[ f_j,e_k
	\right]=\delta _{jk}h_k
\end{equation}
The summation convention does not apply to Eqs.\ (\ref{eq14}) and
$A$ is the Cartan matrix of $F_4$:
%
%
\begin{equation}
\label{eq15}
	A=
	\left(
	\begin{array}{cccc}
		2 & -1 & 0 & 0\\
		-1 & 2 & -1 & 0\\
		0 & -2 & 2 & -1\\
		0 & 0 & -1 & 2
	\end{array}
	\right)
\end{equation}
A solution of Eqs.\ (\ref{eq14}) for the simple generators $e_i$
is as follows:
\begin{eqnarray}
	e_1 &=& D_{7}^5+D_{9}^6+D_{10}^{8}\label{eq16}\\
	e_2 &=& D_{5}^3+D_{6}^4+D_{11}^{10}\label{eq17}\\
	e_3 &=& 2^{-\frac{1}{2}}\left( D_{3}^{\bar
	1}+D_{3}^1+D_{4}^2+D_{8}^6+D_{10}^9+D_{12}^{11}
	\right)\label{eq18}\\
	e_4 &=& 2^{-\frac{1}{2}}\left(
	zD_{2}^{\bar1}+z^*D_{2}^1+D_{4}^3+D_{6}^5+D_{9}^7+D_{13}^{12} \right)
\end{eqnarray}
where
%
%
\begin{equation}
\label{eq20}
	z\equiv e^{i\pi /3}
\end{equation}
Except for the renumbering of rows and columns and a different
choice of phases, my expressions for $e_1$ and $e_2$ are precisely
the same as those given by Patera\cite{patera} and Ekins and
Cornwell\cite{ekins}. However my expressions for $e_3$ and $e_4$
differ from the corresponding expressions of those authors. It
would seem that they resolved some of the arbitrariness in the
solution by demanding that it be real; I require that it display
the antisymmetry across the antidiagonal corresponding to the fact
that we have an orthogonal representation.

In accordance with my labeling of generators of $F_4$ in the $B_4$
basis in terms of the adjoint and the spinor of $B_4$ I have that
the above simple generators $e_i$ should be labeled as follows:
%
%
\begin{equation}
\label{eq20?}
	\begin{array}{r@{=}l@{\rightarrow}l@{=}l}
		\alpha _1 & u_2-u_3 & e_1 & B_{2}^3\\
		\alpha _2 & u_3-u_4 & e_2 & B_{3}^4\\
		\alpha _3 & u_4 & e_3 & B_{4}^0\\
		\alpha _4 & -\frac{1}{2}\left( u_1+u_2+u_3+u_4 \right) & e_4 &
		S^{++++}
	\end{array}
\end{equation}
Next I form commutators of the simple generators and obtain level
one generators
\begin{eqnarray}
	\alpha _1+\alpha _2 &=& u_2-u_4\rightarrow B_{2}^4=\left[
	B_{2}^3,B_{3}^4 \right]=D_{7}^3+D_{9}^4-D_{11}^8\nonumber \\
	\alpha _2+\alpha _3 &=& u_3\rightarrow B_{3}^0=\left[
	B_{3}^4,B_{4}^0 \right]=2^{-\frac{1}{2}}\left(
	D_{5}^{\bar1}+D_{5}^1+D_{6}^2-D_{8}^4+D_{11}^9-D_{12}^{10} \right)\nonumber \\
	\alpha _3+\alpha _4 &=& -\frac{1}{2}(u_1+u_2+u_3-u_4)\rightarrow
	S^{+++-}=\left[ B_{4}^0,S^{++++}
	\right]\sqrt{2}\nonumber \\
	 &=& -2^{-\frac{1}{2}}\left(
	z^*D_{4}^{\bar1}+zD_{4}^1+D_{3}^{\bar2}-D_{8}^5-D_{10}^{7}+D_{13} ^{11}
	\right)\label{eq21}
\end{eqnarray}
level two generators
\begin{eqnarray}
	\alpha _1+\alpha _2+\alpha _3 &=& u_2\rightarrow B_{2}^0=\left[
	B_{2}^3,B_{3}^0 \right]=2^{-\frac{1}{2}}\left(
	D_{7}^{\bar1}+D_{7}^1+D_{9}^2-D_{10}^4-D_{11}^6+D_{12}^8 \right)\nonumber \\
	\alpha _2+\alpha _3+\alpha _4 &=& -\frac{1}{2}\left(
	u_1+u_2-u_3+u_4 \right)\rightarrow S^{++-+}\nonumber \\
	 &=& \left[
	B_{3}^0,S^{++++} \right]\sqrt{2}=-2^{-\frac{1}{2}}\left(
	z^*D_{6}^{\bar1}+zD_{6}^1+D_{5}^{\bar2}+D_{8}^3-D_{11}^7-D_{13}^{10}
\right)\nonumber \\
	\alpha _2+2\alpha _3 &=& u_3+u_4\rightarrow B_{3}^{\bar4}=\left[
	B_{4}^0,B_{3}^0 \right]=D_{5}^{\bar3}+D_{8}^2+D_{12}^9\label{eq22}
\end{eqnarray}
level three generators
\begin{eqnarray}
	\alpha _1+\alpha _2+\alpha _3+\alpha _4 &=& -\frac{1}{2}\left(
	u_1-u_2+u_3+u_4 \right)\rightarrow S^{+-++}=\left[
	B_{2}^0,S^{++++} \right]\sqrt{2}\nonumber \\
	 &=& -2^{-\frac{1}{2}}\left(
	 z^*D_{9}^{\bar1}+zD_{9}^1+D_{7}^{\bar2}+D_{10}^3+D_{11}^5+D_{13}^8
\right)\nonumber \\
	 \alpha _1+\alpha _2+2\alpha _3 &=& u_2+u_4\rightarrow
	 B_{2}^{\bar4}=\left[ B_{4}^0,B_{2}^0
	 \right]=D_{7}^{\bar3}-D_{12}^6+D_{10}^2\nonumber \\
	 \alpha _2+2\alpha _3+\alpha _4 &=& -\frac{1}{2}\left(
	 u_1+u_2-u_3-u_4 \right)\rightarrow S^{++--}=\left[ B_{3}^0,
	 S^{+++-} \right]\sqrt{2}\nonumber \\
	  &=& -2^{-\frac{1}{2}}\left(
	  zD_{8}^{\bar1}+z^*D_{8}^1-D_{6}^{\bar3}-D_{5}^{\bar4}+D_{12}^7-D_{13}^9
	  \right)\label{eq23}
\end{eqnarray}
level four generators
\begin{eqnarray}
	\alpha _1+\alpha _2+2\alpha _3+\alpha _4 &=&
	-\frac{1}{2}\left( u_1-u_2+u_3-u_4 \right)\rightarrow
	S^{+-+-}=\left[ B_{2}^0,S^{+++-} \right]\sqrt{2}\nonumber \\
	 &=& -2^{-\frac{1}{2}}\left(
	 zD_{10}^{\bar1}+z^*D_{10}^1-D_{7}^{\bar4}-D_{9}^{\bar3}-D_{12}^5+D_{13}^6
\right)\nonumber \\
	 \alpha _1+2\alpha _2+2\alpha _3 &=& u_2+u_3\rightarrow
	 B_{2}^{\bar3}=\left[ B_{3}^{\bar4},B_{2}^4
	 \right]=D_{7}^{\bar5}+D_{12}^4+D_{11}^2\nonumber \\
	 \alpha _2+2\alpha _3+2\alpha _4 &=& -u_1-u_2\rightarrow
	 B_{\bar1}^2=\left[ S^{++-+},S^{+++-}
	 \right]=D_{4}^{\bar6}+D_{8}^{\bar2}+D_{13}^7\label{eq24}
\end{eqnarray}
level five generators
\begin{eqnarray}
	\alpha _1+\alpha _2+2\alpha _3+2\alpha _4 &=& -u_1-u_3\rightarrow
	B_{\bar 1}^3=\left[ B_{\bar1}^2,B_{2}^3
	\right]=D_{9}^{\bar4}-D_{10}^{\bar2}+D_{13}^5\nonumber \\
	\alpha _1+2\alpha _2+2\alpha _3+\alpha _4 &=& -\frac{1}{2}\left(
	u_1-u_2-u_3+u_4 \right)\rightarrow S^{+--+}=\left[
	B_{3}^0,S^{+-++} \right]\sqrt{2}\nonumber \\
	 &=& 2^{-\frac{1}{2}}\left(
	 zD_{11}^{\bar1}+z^*D_{11}^1-D_{7}^{\bar6}-D_{9}^{\bar5}+D_{12}^3-D_{13}^4
	 \right)\label{eq25}
\end{eqnarray}
level six generators
\begin{eqnarray}
	\alpha _1+2\alpha _2+2\alpha _3+2\alpha _4 &=&
	-u_1-u_4\rightarrow B_{\bar1}^4=\left[ B_{\bar1}^2,B_{2}^4
	\right]=D_{6}^{\bar9}+D_{11}^{\bar2}+D_{13}^3\nonumber \\
	\alpha _1+2\alpha _2+3\alpha _3+\alpha _4 &=& -\frac{1}{2}\left(
	u_1-u_2-u_3-u_4 \right)\rightarrow S^{+---}=-\left[ B_{4}^0,S^{+--+}
	\right]\sqrt{2}\nonumber \\
	 &=& 2^{\frac{1}{2}}\left(
z^*D_{12}^{\bar1}+zD_{12}^1+D_{7}^{\bar8}+D_{10}^{\bar5}-D_{11}^{\bar3}-D_{13}^2
	 \right)\label{eq26}
\end{eqnarray}
one level seven generator
\begin{eqnarray}
	\alpha _1+2\alpha _2+3\alpha _3+2\alpha _4 &=& -u_1\rightarrow
	B_{\bar1}^0=-\left[ S^{+---},S^{++++} \right]\sqrt{2}\nonumber \\
	 &=& 2^{\frac{1}{2}}\left(
D_{13}^{\bar1}+D_{13}^1+D_{9}^{\bar8}+D_{10}^{\bar6}-D_{11}^{\bar4}-D_{12}^{\bar2}
	 \right)\label{eq27}
\end{eqnarray}
one level eight generator
%
%
\begin{equation}
\label{eq28}
	\alpha _1+2\alpha _2+4\alpha _3+2\alpha _4=-u_1+u_4\rightarrow
	B_{4}^1=\left[ B_{\bar1}^0,B_{4}^0
	\right]=-D_{10}^{\bar8}+D_{12}^{\bar4}-D_{13}^{\bar3}
\end{equation}
one level nine generator
%
%
\begin{equation}
\label{eq29}
	\alpha _1+3\alpha _2+4\alpha _3+2\alpha _4=-u_1+u_3\rightarrow
	B_{3}^1=\left[ B_{\bar1}^0,B_{3}^0
	\right]=-D_{11}^{\bar8}+D_{12}^{\bar6}-D_{13}^{\bar5}
\end{equation}
and one level ten generator
%
%
\begin{equation}
\label{eq30}
	2\alpha _1+3\alpha _2+4\alpha _3+2\alpha _4=-u_1+u_2\rightarrow
	B_{2}^1=\left[ B_{\bar1}^0, B_{2}^0
	\right]=D_{10}^{\overline{11}}+D_{12}^{\bar9}-D_{13}^{\bar7}
\end{equation}
Note that the root corresponding to the highest level, Eq.\
(\ref{eq30}), is precisely the negative of $\alpha _0$, where
$\alpha _0$ is the extra root added to the Dynkin diagram of $F_4$ to
produce the extended Dynkin diagram.

In addition to the above 24 $e$-type generators, Eqs.\
(\ref{eq16})--(\ref{eq30}), I have 24 $f$-type generators obtained
by taking the hermitian conjugate of the above. Thus corresponding
to the expressions above for the simple (level zero) lowering
generators $e_i$ I have
\begin{eqnarray}
	f_1 &=&
	e_1^\dagger=B_{2}^{3^\dagger}=B_{3}^2=D_{5}^7+D_{6}^9+D_{8}^{10}\nonumber \\
	f_2 &=&
e_{2}^{\dagger}=B_{3}^{4^{\dagger}}=B_{4}^3=D_{3}^5+D_{4}^6+D_{10}^{11}\nonumber \\
	f_3 &=&
	e_3^\dagger=B_{4}^{0^{\dagger}}=B_{0}^4=2^{\frac{1}{2}}\left(
	D_{\bar1}^3+D_{1}^3+D_{2}^4+D_{6}^8+D_{9}^{10}+D_{11}^{12} \right)\nonumber \\
	f_4 &=&
	e_{4}^\dagger=S^{++++^\dagger}=S^{----}=2^{\frac{1}{2}}\left(
	z^*D_{\bar1}^2+zD_{1}^2+D_{3}^4+D_{5}^6+D_{7}^9+D_{12}^{13}
	\right)\label{eq31}
\end{eqnarray}
and so on for the generators in higher levels.

Moreover, for the hermitian Cartan generators I have that the
Chevalley and $B_4$ bases are related as follows:
\begin{eqnarray}
	h_1 &=& \left[ f_1,e_1 \right]=\left[ B_{3}^2,B_{2}^3
	\right]=B_{3}^3-B_{2}^2\nonumber \\
	h_2 &=& \left[ f_2,e_2 \right]=\left[ B_{4}^3,B_{3}^4
	\right]=B_{4}^4-B_{3}^3\nonumber \\
	h_3 &=& \left[ f_3,e_3 \right]=\left[ B_{0}^4,B_{4}^0
	\right]=-B_{4}^4\nonumber \\
	h_4 &=& \left[ f_4,e_4 \right]=\left[ S^{----},S^{++++}
	\right]=\frac{1}{2}\left( B_{1}^1+B_{2}^2+B_{3}^3+B_{4}^4
	\right)\label{eq32}
\end{eqnarray}
or, solving above for the $B_{\alpha }^{\alpha }$ and using Eqs.\
(\ref{eq10})--(\ref{eq13}),
\begin{eqnarray}
	-B_{1}^1 &=& h_1+2h_2+3h_3+2h_4\nonumber \\
	 &=& \frac{1}{2}\left(
D_{2}^2+D_{4}^4+D_{6}^6+D_{8}^8+D_{9}^9+D_{10}^{10}+D_{11}^{11}+D_{12}^{12}+2D_{13}^{13} \right)\nonumber \\
	-B_{2}^2 &=& h_1+h_2+h_3=\frac{1}{2}\left(
D_{2}^2+D_{4}^4+D_{6}^6-2D_{7}^7+D_{8}^8-D_{9}^9-D_{10}^{10}-D_{11}^{11}-D_{12}^{12} \right)\nonumber \\
	-B_{3}^3 &=& h_2+h_3=\frac{1}{2}\left(
D_{2}^2+D_{4}^4-2D_{5}^5-D_{6}^6-D_{8}^8+D_{9}^9+D_{10}^{10}-D_{11}^{11}-D_{12}^{12}\right)\nonumber \\
	-B_{4}^4 &=& h_3=\frac{1}{2}\left(
D_{2}^2-2D_{3}^3-D_{4}^4+D_{6}^6-D_{8}^8+D_{9}^9-D_{10}^{10}+D_{11}^{11}-D_{12}^{12}
	\right)\label{eq33}
\end{eqnarray}

\section*{The projected generators $\tilde D_{a}^b$}

Now since the 26 is the defining representation of $D_{13}$, the
results above expressing the generators of $F_4$ in the
26-dimensional representation in terms of the generators of
$D_{13}$ in the 26-dimensional representation, can be interpreted
as giving the generators of $F_4$ in terms of those of $D_{13}$ in
any representation. Now then the $\tilde D_{a}^b$, the generators
of $D_{13}$ projected into $F_4$, are given by inverting the above
equations.

Thus the 13 Cartan generators of $D_{13}$ projected into $F_4$ are
given by inverting Eqs. (\ref{eq33}):
\begin{eqnarray}
	\tilde D_{1}^1 &=& 0\nonumber \\
	\tilde D_{2}^2 &=& -\frac{1}{6}\left( B_{1}^1+B_{2}^2+B_{3}^3+B_{4}^4
\right)\nonumber \\
	\tilde D_{3}^3 &=& \frac{1}{3}B_{4}^4\nonumber \\
	\tilde D_{4}^4 &=& -\frac{1}{6}\left(
	B_{1}^1+B_{2}^2+B_{3}^3-B_{4}^4 \right)\nonumber \\
	\tilde D_{5}^5 &=& \frac{1}{3}B_{3}^3\nonumber \\
	\tilde D_{6}^6 &=& -\frac{1}{6}\left(
	B_{1}^1+B_{2}^2-B_{3}^3+B_{4}^4 \right)\nonumber \\
	\tilde
	D_{7}^7 &=& \frac{1}{3}B_{2}^2\nonumber \\
	\tilde D_{8}^8 &=& -\frac{1}{6}\left(
	B_{1}^1+B_{2}^2-B_{3}^3-B_{4}^4 \right)\nonumber \\
	\tilde D_{9}^9 &=& -\frac{1}{6}\left(
	B_{1}^1-B_{2}^2+B_{3}^3+B_{4}^4 \right)\nonumber \\
	\tilde D_{10}^{10} &=& -\frac{1}{6}\left(
	B_{1}^1-B_{2}^2+B_{3}^3-B_{4}^4 \right)\nonumber \\
	\tilde D_{11}^{11} &=& -\frac{1}{6}\left(
	B_{1}^1-B_{2}^2-B_{3}^3+B_{4}^4 \right)\nonumber \\
	\tilde D_{12}^{12} &=& -\frac{1}{6}\left(
	B_{1}^1-B_{2}^2-B_{3}^3-B_{4}^4 \right)\nonumber \\
	\tilde D_{13}^{13} &=& -\frac{1}{3}B_{1}^1\label{eq34}
\end{eqnarray}
Perhaps an explanation of how Eq.\ (\ref{eq34}) is obtained is in
order. Equations (\ref{eq33}) are four equations for four
$B_{\alpha }^{\alpha }$ in terms of thirteen $D_{a}^a$ (no
summations). In addition there are nine more equations for
appropriate components of the $\bf 273$-plet involving these same
thirteen $D_{a}^a$. This total of 13 equations can be written as
follows
%
%
\begin{equation}
\label{eq35}
	b_A=U_{AB}d_B
\end{equation}
where $1\leq A,B\leq 13$, where $d_B\equiv D_{B}^B$ (no
summation), $b_A\equiv B_{A}^A$ (no summation) for $A=1,2,3,4$,
and $b_A$ for $5\leq A\leq 13$ refers to components of the $\bf
273$-plet. Inversion of Eq.\ (\ref{eq35}) is achieved by
%
%
\begin{equation}
\label{eq36}
	d_A=U^{-1}_{AB}b_B
\end{equation}
where the inverse of the $13\times13$ matrix $U$ is given by
%
%
\begin{equation}
\label{eq37}
	U^{-1}=\frac{1}{3}U^\dagger
\end{equation}
where the factor $\frac{1}{3}$ accounts for the difference in the
normalization of the $d_A$ and $b_A$. Finally the projected
$\tilde d_A$ are obtained by setting in Eq.\ (\ref{eq36}) $b_A=0$
for $5\leq A\leq13$.

By proceeding in the same fashion I obtain the 156 generators
$\tilde D_{a}^b$ with $a>b$ by inverting the 24 $e$-type equations
with the result:\\
for the 24 $\tilde D_{13}^b$ with $13>b$:
\begin{eqnarray}
	\tilde D_{13}^{\overline{12}} &=& \tilde
	D_{13}^{\overline{11}}=\tilde D_{13}^{\overline{10}}=\tilde
	D_{13}^{\bar9}=\tilde D_{13}^{\bar8}=\tilde D_{13}^{\bar6}=\tilde
	D_{13}^{\bar4}=\tilde D_{13}^{\bar2}=0,\nonumber \\
	\tilde D_{13}^{\bar7} &=& -\frac{1}{3}B_{2}^1,\ \tilde
	D_{13}^{\bar5}=-\frac{1}{3}B_{3}^1,\ \tilde
	D_{13}^{\bar3}=-\frac{1}{3}B_{4}^1,\nonumber \\
	\tilde D_{13}^{\bar1} &=&
	\tilde D_{13}^1=
	-\frac{1}{3\sqrt{2}}B_{0}^1,\ \tilde
	D_{13}^2=-\frac{1}{3\sqrt{2}}S^{+---},\nonumber \\
	\tilde D_{13}^3 &=& \frac{1}{3}B_{\bar1}^4,\ \tilde
	D_{13}^4=-\frac{1}{3\sqrt{2}}S^{+--+},\ \tilde
	D_{13}^5=\frac{1}{3}B_{\bar1}^3,\nonumber \\
	\tilde D_{13}^6 &=& -\frac{1}{3\sqrt{2}}S^{+-+-},\ \tilde
	D_{13}^7=\frac{1}{3}B_{\bar1}^2,\ \tilde
	D_{13}^8=-\frac{1}{3\sqrt{2}}S^{+-++},\nonumber \\
	\tilde D_{13}^9 &=& \frac{1}{3\sqrt{2}}S^{++--},\ \tilde
	D_{13}^{10}=\frac{1}{3\sqrt{2}}S^{++-+},\nonumber \\
	\tilde D_{13}^{11} &=& -\frac{1}{3\sqrt{2}}S^{+++-},\ \tilde
	D_{13}^{12}=\frac{1}{3\sqrt{2}}S^{++++}\label{eq38}
\end{eqnarray}
for the 22 $\tilde D_{12}^b$ with $12>|b|$:
\begin{eqnarray}
	\tilde D_{12}^{\overline{11}} &=& \tilde
	D_{12}^{\overline{10}}=\tilde D_{12}^{\bar8}=\tilde
	D_{12}^{\bar7}=\tilde D_{12}^{\bar5}=\tilde D_{12}^{\bar3}=\tilde
	D_{12}^{2}=0,\nonumber \\
	\tilde D_{12}^{\bar9} &=& \frac{1}{3}B_{2}^1,\ \tilde
	D_{12}^{\bar6}=\frac{1}{3}B_{3}^1,\ \tilde
	D_{12}^{\bar4}=\frac{1}{3}B_{4}^1\nonumber \\
	\tilde D_{12}^{\bar2} &=& \frac{1}{3\sqrt{2}}B_{0}^1,\ \tilde
	D_{12}^{\bar1}=\frac{z}{3\sqrt{2}}S^{+---},\ \tilde
	D_{12}^1=\frac{z^*}{3\sqrt{2}}S^{+---},\nonumber \\
	\tilde D_{12}^3 &=& \frac{1}{3\sqrt{2}}S^{+--+},\ \tilde
	D_{12}^4=\frac{1}{3}B_{2}^{\bar3},\ \tilde
	D_{12}^5=\frac{1}{3\sqrt{2}}S^{+-+-},\nonumber \\
	\tilde D_{12}^6 &=& \frac{1}{3}B_{4}^{\bar2},\ \tilde
	D_{12}^7=-\frac{1}{3\sqrt{2}}S^{++--},\ \tilde
	D_{12}^8=\frac{1}{3\sqrt{2}}B_{2}^0,\nonumber \\
	\tilde D_{12}^9 &=& \frac{1}{3}B_{3}^{\bar4},\ \tilde
	D_{12}^{10}=\frac{1}{3\sqrt{2}}B_{0}^{\bar3},\ \tilde
	D_{12}^{11}=\frac{1}{3\sqrt{2}}B_{4}^0\label{eq39}
\end{eqnarray}
for the 20 $\tilde D_{11}^b$ with $11>|b|$:
\begin{eqnarray}
	\tilde D_{11}^{\bar9} &=& \tilde D_{11}^{\bar7}=\tilde
	D_{11}^{\bar6}=\tilde D_{11}^{\bar5}=\tilde D_{11}^{3}=\tilde
	D_{11}^4=0,\nonumber \\
	\tilde D_{11}^{\overline{10}} &=& -\frac{1}{3}B_{2}^1,\ \tilde
	D_{11}^{\overline8}=-\frac{1}{3}B_{3}^1,\ \tilde
	D_{11}^{\overline4}=\frac{1}{3\sqrt{2}}B_{0}^1,\nonumber \\
	\tilde D_{11}^{\overline3} &=& -\frac{1}{3\sqrt{2}}S^{+---},\
	\tilde D_{11}^{\overline2}=\frac{1}{3}B_{\overline1}^4,\ \tilde
	D_{11}^{\overline1}=\frac{z^*}{3\sqrt{2}}S^{+--+},\nonumber \\
	\tilde D_{11}^1 &=& \frac{z}{3\sqrt{2}}S^{+--+},\ \tilde
	D_{11}^2=\frac{1}{3}B_{2}^{\bar3},\ \tilde
	D_{11}^5=-\frac{1}{3\sqrt{2}}S^{+-++},\nonumber \\
	\tilde D_{11}^6 &=& \frac{1}{3\sqrt{2}}B_{0}^{\bar2},\ \tilde
	D_{11}^7=\frac{1}{3\sqrt{2}}S^{++-+},\ \tilde
	D_{11}^8=-\frac{1}{3}B_2^4,\nonumber \\
	\tilde D_{11}^9 &=& \frac{1}{3\sqrt{2}}B_2^0,\ \tilde
	D_{11}^{10}=\frac{1}{3}B_3^4\label{eq40}
\end{eqnarray}
for the 18 $\tilde D_{10}^b$ with $10>|b|$:
\begin{eqnarray}
	\tilde D_{10}^{\bar9} &=& \tilde D_{10}^{\bar7}=\tilde
	D_{10}^{\bar4}=\tilde D_{10}^{\bar3}=\tilde D_{10}^{5}=\tilde
	D_{10}^{6}=0,\nonumber \\
	\tilde D_{10}^{\bar8} &=& -\frac{1}{3}B_4^1,\ \tilde
	D_{10}^{\bar5}=\frac{1}{3\sqrt{2}}S^{+---},\ \tilde
	D_{10}^{\bar2}=-\frac{1}{3}B_{\bar1}^3\nonumber \\
	\tilde D_{10}^{\bar1} &=& -\frac{z^*}{3\sqrt{2}}S^{+-+-},\ \tilde
	D_{10}^1=-\frac{z}{3\sqrt{2}}S^{+-+-},\ \tilde
	D_{10}^2=\frac{1}{3}B_2^{\bar4},\nonumber \\
	\tilde D_{10}^3 &=& -\frac{1}{3\sqrt{2}}S^{+-++},\ \tilde
	D_{10}^4=\frac{1}{3\sqrt{2}}B_{0}^{\bar2},\ \tilde
	D_{10}^7=\frac{1}{3\sqrt{2}}S^{+++-},\nonumber \\
	\tilde D_{10}^8 &=& \frac{1}{3}B_2^3,\ \tilde
	D_{10}^9=\frac{1}{3\sqrt{2}}B_4^0\label{eq41}
\end{eqnarray}
for the 16 $\tilde D_{9}^b$ with $9>|b|$:
\begin{eqnarray}
	\tilde D_9^{\bar7} &=& \tilde D_9^{\bar2}=\tilde D_9^3=\tilde
	D_9^5=\tilde D_9^8=0,\nonumber \\
	\tilde D_9^{\bar8} &=& -\frac{1}{3\sqrt{2}}B_0^1,\ \tilde
	D_9^{\bar6}=\frac{1}{3}B_{\bar4}^1,\ \tilde
	D_9^{\bar5}=-\frac{1}{3\sqrt{2}}S^{+--+},\nonumber \\
	\tilde D_9^{\bar4} &=& \frac{1}{3}B_{\bar1}^3,\ \tilde
	D_9^{\bar3}=\frac{1}{3\sqrt{2}}S^{+-+-},\ \tilde
	D_9^{\bar1}=-\frac{z}{3\sqrt{2}}S^{+-++},\nonumber \\
	\tilde D_9^1 &=& -\frac{z^*}{3\sqrt{2}}S^{+-++},\ \tilde
	D_9^2=\frac{1}{3\sqrt{2}}B_2^0,\ \tilde D_9^4=\frac{1}{3}B_2^4,\nonumber \\
	\tilde D_9^6 &=& \frac{1}{3}B_2^3,\ \tilde
	D_9^7=\frac{1}{3\sqrt{2}}S^{++++}\label{eq42}
\end{eqnarray}
for the 14 $\tilde D_8^b$ with $8>|b|$:
\begin{eqnarray}
	\tilde D_8^{\bar6} &=& \tilde D_8^{\bar5}=\tilde
	D_8^{\bar4}=\tilde D_8^{\bar3}=\tilde D_8^7=0,\nonumber \\
	\tilde D_8^{\bar7} &=& -\frac{1}{3\sqrt{2}}S^{+---},\ \tilde
	D_8^{\bar2}=\frac{1}{3}B_{\bar1}^2,\ \tilde
	D_8^{\bar1}=-\frac{z^*}{3\sqrt{2}}S^{++--},\nonumber \\
	\tilde D_8^1 &=& -\frac{z}{3\sqrt{2}}S^{++--},\ \tilde
	D_8^2=\frac{1}{3}B_3^{\bar4},\ \tilde
	D_8^3=-\frac{1}{3\sqrt{2}}S^{++-+},\nonumber \\
	\tilde D_8^4 &=& -\frac{1}{3\sqrt{2}}B_3^0,\ \tilde
	D_8^5=\frac{1}{3\sqrt{2}}S^{+++-},\ \tilde
	D_8^6=\frac{1}{3\sqrt{2}}B_4^0\label{eq43}
\end{eqnarray}
for the 12 $\tilde D_7^b$ with $7>|b|$:
\begin{eqnarray}
	\tilde D_7^2 &=& \tilde D_7^4=\tilde D_7^6=0,\ \tilde
	D_7^{\bar6}=-\frac{1}{3\sqrt{2}}S^{+--+},\ \tilde
	D_7^{\bar5}=\frac{1}{3}B_2^{\bar3},\nonumber \\
	\tilde D_7^{\bar4} &=& \frac{1}{3\sqrt{2}}S^{+-+-},\ \tilde
	D_7^{\bar3}=\frac{1}{3}B_2^{\bar4},\ \tilde
	D_7^{\bar2}=-\frac{1}{3\sqrt{2}}S^{+-++},\nonumber \\
	\tilde D_7^{\bar1} &=& \tilde D_7^{1}=\frac{1}{3\sqrt{2}}B_2^0,\
	\tilde D_7^3=\frac{1}{3}B_2^4,\ \tilde
	D_7^5=\frac{1}{3}B_2^3\label{eq44}
\end{eqnarray}
for the 10 $\tilde D_6^b$ with $6>|b|$:
\begin{eqnarray}
	\tilde D_6^{\bar5} &=& \tilde D_6^{\bar2}=\tilde D_6^3=0,\
	\tilde D_6^{\bar4}=-\frac{1}{3}B_{\bar1}^2,\
	\tilde D_6^{\bar3}=\frac{1}{3\sqrt{2}}S^{++--},\nonumber \\
	\tilde D_6^{\bar1} &=& -\frac{z}{3\sqrt{2}}S^{++-+},\
	\tilde D_6^1=-\frac{z^*}{3\sqrt{2}}S^{++-+},\
	\tilde D_6^2=\frac{1}{3\sqrt{2}}B_3^0,\nonumber \\
	\tilde D_6^4 &=& \frac{1}{3}B_3^4,\
	\tilde D_6^5=\frac{1}{3\sqrt{2}}S\mbox{++++}\label{eq45}
\end{eqnarray}
for the 8 $\tilde D_5^b$ with $5>|b|$:
\begin{eqnarray}
	\tilde D_5^2 &=& \tilde D_5^4=0,\ \tilde
	D_5^{\bar4}=\frac{1}{3\sqrt{2}}S^{++--},\ \tilde
	D_5^{\bar3}=\frac{1}{3}B_3^{\bar4},\nonumber \\
	\tilde D_5^{\bar2} &=& -\frac{1}{3\sqrt{2}}S^{++-+},\ \tilde
	D_5^{\bar1}=\tilde D_5^1=\frac{1}{3\sqrt{2}}B_3^0,\nonumber \\
	\tilde D_5^3 &=& \frac{1}{3}B_3^4\label{eq46}
\end{eqnarray}
for the 6 $D_4^b$ with $4>|b|$:
\begin{eqnarray}
	\tilde D_4^{\bar3} &=& \tilde D_4^{\bar2}=0,\ \tilde
	D_4^{\bar1}=-\frac{z}{3\sqrt{2}}S^{+++-},\ \tilde
	D_4^1=-\frac{z^*}{3\sqrt{2}}S^{+++-},\nonumber \\
	\tilde D_4^2 &=& \frac{1}{3\sqrt{2}}B_4^0,\ \tilde
	D_4^3=\frac{1}{3\sqrt{2}}S^{++++}\label{eq47}
\end{eqnarray}
for the 4 $\tilde D_3^b$ with $3>|b|$:
%
%
\begin{equation}
\label{eq48}
	\tilde D_3^2=0,\ \tilde
	D_3^{\bar2}=-\frac{1}{3\sqrt{2}}S^{+++-},\ \tilde
	D_3^{\bar1}=\tilde D_3^1=\frac{1}{3\sqrt{2}}B_4^0
\end{equation}
and finally for the two $\tilde D_2^b$ with $2>|b|$:
%
%
\begin{equation}
\label{eq49}
	\tilde D_2^{\bar1}=\frac{z^*}{3\sqrt{2}}S^{++++},\ \tilde
	D_2^1=\frac{z}{3\sqrt{2}}S^{++++}
\end{equation}
Lastly the 156 $\tilde D_a^b$ with $a<b$ are obtained from the
results above by hermitian conjugation:
%
%
\begin{equation}
\label{eq50}
	\tilde D_b^a=\tilde D_a^{b^\dagger},\ B_b^a=B_a^{b^\dagger},\
	S^{\overline{pqrs}}=S^{pqrs^\dagger}
\end{equation}
This completes the calcultion of the Casimir operators of $F_4$.

\section*{Conclusion}

I conclude with two remarks,
\begin{enumerate}
	\item For $k=2$
	the result of inserting the explicit formulas for the projected $\tilde
	D$, Eqs.\ (\ref{eq34}), (\ref{eq38}--\ref{eq50}), into Eq.\
	(\ref{eq6}) can be simplified into the following formula for the
	quadratic Casimir operator of $F_4$:
	%
	%
	\begin{equation}
	\label{eq51}
		{\cal C}_2(F_4)=\tilde D_a^b\tilde D_b^a=\frac{1}{3}B_\alpha
		^\beta B_\beta ^\alpha +\frac{2}{3}S^{pqrs}S^{\overline{pqrs}}
	\end{equation}
	and I remind the reader that the various subscripts are
	summed over the following range: $-13\leq a,b\leq13$ (zero
	excluded); $-4\leq \alpha ,\beta \leq4$ (zero included);
	$p,q,r,s=\pm$.

	The general form of this result for the quadratic Casimir of
	$F_4$ in the $B_4$ basis was to be expected since the two pieces
	in Eq.\ (\ref{eq51}) are the only quadratic invariants of the
	subgroup $B_4$ that can be formed out of the adjoint $\bf 36$ and
	the spinor $\bf16$ of $B_4$. Thus this result can be viewed as a
	test of the formalism.

	\item Recall that according to Eq.\ (\ref{eq4}) the independent Casimirs
	are of degree $k=2s,\ 1\leq s\leq13$. Now consider the {\em
	Cartan\/} part of the Casimirs. If I denote the Cartan part of
	${\cal C}_k(F_4)$ by ${\cal K}_k$ then it follows from Eq.\
	(\ref{eq6}) that
	%
	%
	\begin{equation}
	\label{eq52}
		{\cal K}_k=
		\left( \tilde D_a^a \right)^k
	\end{equation}
	Since $\tilde D_{\bar a}^{\bar a}=-\tilde D_a^a$ this is
	manifestly zero for $k=\mbox{odd}$. For $k=\mbox{even}$ Eq.\
	(\ref{eq52}) becomes (where I have set $b_\alpha \equiv B_\alpha
	^\alpha $, no summation)
	\begin{eqnarray}
		&&{\cal K}_k=2\sum_{a=1}^{13}\left( \tilde D_a^a \right)^k =
		2\cdot6^{-k}\left\{\left( b_1+b_2+b_3+b_4 \right)^k+\left( 2b_4
		\right)^k+\left( b_1+b_2+b_3-b_4 \right)^k+\left( 2b_3 \right)^k
		\right.\nonumber \\
		&& +\left( b_1+b_2-b_3+b_4 \right)^k+\left( 2b_2
		\right)^k+\left( b_1+b_2-b_3-b_4 \right)^k+\left(
		b_1-b_2+b_3+b_4 \right)^k\nonumber \\
		&&\left.+\left( b_1-b_2+b_3-b_4 \right)^k+\left( b_1-b_2-b_3+b_4
		\right)^k+\left( b_1-b_2-b_3-b_4 \right)^k+\left( 2b_1
		\right)^k\right\}\label{eq53}
	\end{eqnarray}
	For $k=2$ Eq.\ (\ref{eq53}) gives
	%
	%
	\begin{equation}
	\label{eq54}
		{\cal K}_2=\frac{2}{3}\left( b_2^2+b_2^2+b_3^2+b_4^2 \right)
	\end{equation}
	while for $k=4$ it gives
	%
	%
	\begin{equation}
	\label{eq55}
		{\cal K}_4=3^{-3}\left( b_1^2+b_2^2+b_3^2+b_4^2 \right)^2
	\end{equation}
	This proves that the degree 4 Casimir is not functionally
	independent of the degree 2 Casimir.

	For $k=6$ Eq.\ (\ref{eq53}) gives
	\begin{eqnarray}
		{\cal K}_6 &=& 2^{-2}\cdot 3^{-5}\left\{3\left[
		b_1^6+b_2^6+b_3^6+b_4^6 \right]+5\left[ b_1^4\left(
		b_2^2+b_3^2+b_4^2 \right)+b_2^4\left( b_1^2+b_3^2+b_4^2 \right)
		\right.\right.\nonumber \\
	    && \left.+b_3^4\left( b_1^2+b_2^2+b_4^2 \right)+b_4^4\left(
	    b_1^2+b_2^2+b_3^2 \right) \right]\nonumber \\
	    && \left. +30\left[ b_1^2\left( b_2^2b_3^2+b_3^2b_4^2+b_2^2b_4^2
	    \right)+b_2^2b_3^2b_4^2 \right]\right\}\label{eq56}
	\end{eqnarray}
	which {\em is\/} functionally independent of the degree 2 Casimir
	(were it proportional to the cube of the degree 2 Casimir it
	would have the coefficient of the expression in the first, second
	and third square bracket in the ratio 1:3:6 instead of the 3:5:30
	above).

	Continuing along these lines I find that the degree 8 is
	functionally independent of the degree 2 and 6, while the degree
	10 is dependent:
	%
	%
	\begin{equation}
	\label{eq57}
		{\cal K}_{10}\sim{\cal K}_2\left\{28{\cal K}_2({\cal
		K}_2^3-{\cal K}_6)+3{\cal K}_8\right\}
	\end{equation}
	and lastly the degree 12 is independent of those of lower
	degree. Since all the Casimirs are functions of the four
	quantities $b_\alpha ^2$, $1\leq\alpha \leq4$, I can solve for
	the $b_\alpha ^2$ in terms of the independent Casimirs of degree
	2, 6, 8 and 12, and consequently all Casimirs of higher degree
	are necessarily dependent. This completes the demonstration that
	the independent Casimirs are those of degree equal to the
	exponents of $F_4$ plus one.

\end{enumerate}

	\acknowledgements

	Many useful discussions with Charlie Goebel are gratefully
	acknowledged.

\end{document}